\documentclass[aps,pra,preprint,superscriptaddress,longbibliography,endfloats*]{revtex4-1}
\usepackage{graphicx}
\usepackage{color}
\usepackage{amsmath,amssymb}
\usepackage{bm}
\usepackage{upgreek}
\usepackage{xspace}
\usepackage[colorlinks,urlcolor=blue,citecolor=blue,linkcolor=blue]{hyperref}

\newcommand{\um}{\upmu \text{m}}

\begin{document}

\title{Quantum measurement of a rapidly rotating spin qubit in diamond}

\author{Alexander A.~Wood}
\affiliation{School of Physics, University of Melbourne, Victoria 3010, Australia}
\author{Emmanuel Lilette}
\affiliation{School of Physics, University of Melbourne, Victoria 3010, Australia}
\author{Yaakov Y. Fein}
\affiliation{School of Physics, University of Melbourne, Victoria 3010, Australia}
\author{Nikolas Tomek}
\affiliation{Institut f\"ur Quantenoptik, Universit\"at Ulm, Ulm 89069, Germany}
\author{Liam P. McGuinness}
\affiliation{Institut f\"ur Quantenoptik, Universit\"at Ulm, Ulm 89069, Germany}
\author{Lloyd C. L. Hollenberg}
\affiliation{School of Physics, University of Melbourne, Victoria 3010, Australia}
\author{Robert E. Scholten}
\affiliation{School of Physics, University of Melbourne, Victoria 3010, Australia}
\author{Andy M. Martin}
\email{martinam@unimelb.edu.au}
\affiliation{School of Physics, University of Melbourne, Victoria 3010, Australia}

\date{\today}

\begin{abstract}
A controlled qubit in a rotating frame opens new opportunities to probe fundamental quantum physics, such as geometric phases in physically rotating frames, and can potentially enhance detection of magnetic fields. Realising a single qubit that can be measured and controlled during physical rotation is experimentally challenging. In this work, we demonstrate quantum control of a single nitrogen-vacancy (NV) centre within a diamond rotated at 200,000\,rpm, a rotational period comparable to the NV spin coherence time $T_2$. We stroboscopically image individual NV centres that execute rapid circular motion in addition to rotation, and demonstrate preparation, control and readout of the qubit quantum state with lasers and microwaves. Using spin-echo interferometry of the rotating qubit, we are able to detect modulation of the NV Zeeman shift arising from the rotating NV axis and an external DC magnetic field. Our work establishes single NV qubits in diamond as quantum sensors in the physically rotating frame, and paves the way for the realisation of single-qubit diamond-based rotation sensors.   
\end{abstract}


\maketitle

\section{Introduction}
Profound aspects of fundamental and applied physics can be probed with a single quantum system in a rotating frame, including geometric phase~\cite{berry_quantal_1984, tycko_adiabatic_1987}, effective magnetic fields~\cite{barnett_gyromagnetic_1935} and magic-angle spinning~\cite{andrew_magic_1981}. The nitrogen-vacancy (NV) centre in diamond~\cite{doherty_nitrogen-vacancy_2013, schirhagl_nitrogen-vacancy_2014} is an ideal solid-state qubit to study quantum physics in the rotating frame. In static, non-rotating frames it has demonstrated success as a nanoscale detector of magnetic fields \cite{rondin_magnetometry_2014}, electric fields \cite{dolde_electric-field_2011}, crystal strain \cite{doherty_electronic_2014} and temperature \cite{acosta_temperature_2010} in real-world sensing environments, such as within biological cells \cite{mcguinness_quantum_2011}. An NV centre in a physically rotating diamond brings these extensive quantum sensing modalities to the rotating frame, and is predicted to accumulate a quantum mechanical Berry phase~\cite{maclaurin_measurable_2012, kowarsky_non-abelian_2014}, which underlies the operation of a proposed NV-diamond gyroscope~\cite{ledbetter_gyroscopes_2012}. Rotating diamonds with large ensembles of NV centres have recently been used to explore magnetic pseudo-fields generated in the rotating frame~\cite{wood_magnetic_2017}. This work realises quantum measurement of a single solid-state qubit in a diamond rotating with a period comparable to the spin coherence time, establishing an exemplar system for probing rotational effects at the single quantum level.

Our experiments offer the robust, well-defined environment of bulk diamond in which to study the effects of rotation on a single spin qubit. Hybrid spin-optomechanical systems, consisting of NV-hosting micro- and nanodiamonds in optical~\cite{horowitz_electron_2012, neukirch_multi-dimensional_2015, hoang_electron_2016} and ion traps~\cite{delord_electron_2017}, are of considerable interest for studying quantum superpositions of massive objects~\cite{yin_large_2013, scala_matter-wave_2013} and could potentially also be used to explore rotation of qubits. However, free rotation of the qubit in trapped-nanodiamond experiments is difficult to control, as it depends on the trapping (or measurement) laser and geometry of the nanodiamond itself~\cite{hoang_torsional_2016,delord_electron_2017}, scrambling the orientation of the NV axis. The typical coherence times of NVs hosted in nanodiamonds $(T_2 < 1\,\upmu\text{s})$ require extremely rapid rotations to study rotational physics in a genuinely quantum setting. We demonstrate control and optical state readout of a single NV qubit during rapid, well-controlled physical rotation, and perform coherent quantum measurements over a significant fraction of the rotation period. We use quantum measurement in the rotating frame to measure an effect unique to a rotating NV centre -- time-varying Zeeman shifts arising from stationary-frame DC magnetic fields. Our work opens the study of frame-dependent effects to quantum sensing at the single-spin level with a robust, versatile quantum sensor.      

\section{Results}
\subsection{Experimental setup}
\begin{figure*}
	\centering
		\includegraphics{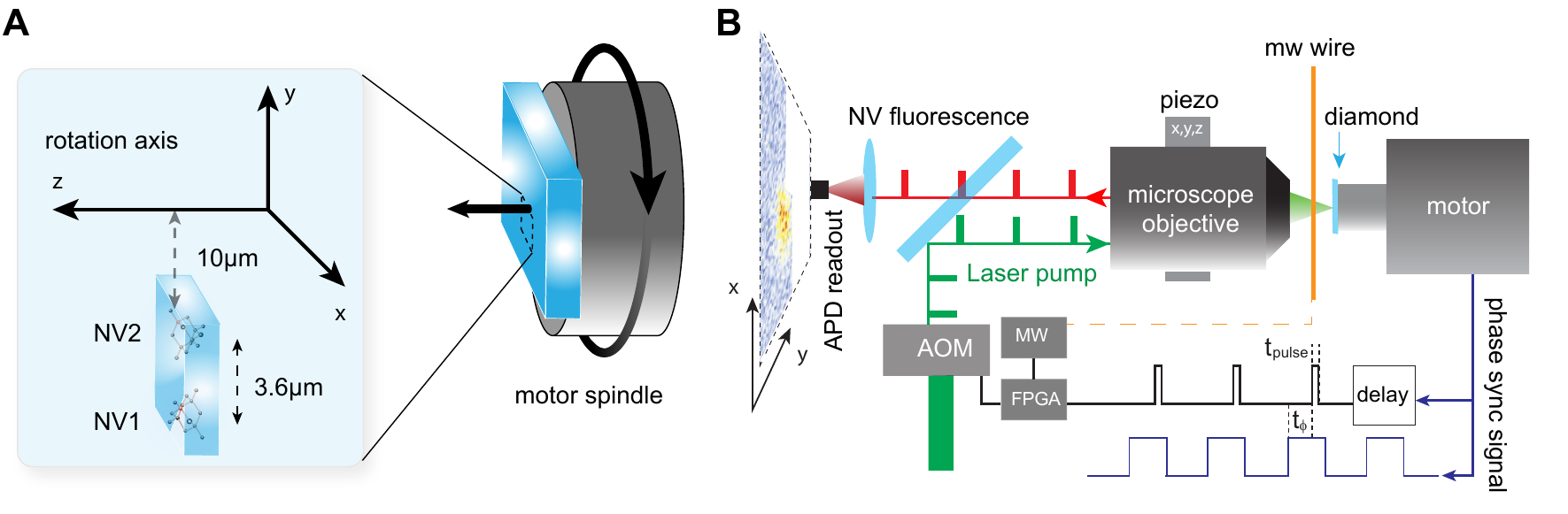}
	\caption{{\bf Experimental schematic and setup.} ({\bf A}) A diamond containing single NV centres is mounted to the spindle of an electric motor that rotates at 3.33\,kHz. Two NV centres, located $3\um$ below the diamond surface, separated by $3.6\um$ and located $10\um$ from the centre of the rotation axis ($z$) are considered in this work. ({\bf B}) NV centres are optically prepared and addressed by a scanning confocal microscope: $532$\,nm excitation light is pulsed for a duration $t_\text{pulse}$ synchronously with the motor rotation, and an avalanche photodiode (APD) collects the emitted photons from the NV centre. A time delay $t_\phi$ between the motor synchronisation edge and the optical pulse controls the instantaneous angle of the diamond imaged, and a wire located above the surface of the diamond is used to apply microwave pulses for state manipulation. Current-carrying coils (not shown) are used to apply magnetic fields along the $x$, $y$ and $z$ axes.}
	\label{fig:apparatus}
\end{figure*}

A schematic of our experiment is depicted in Fig. \ref{fig:apparatus}. The diamond sample contains well-separated single NV centres $3\,\upmu$m below the surface, with the $\langle 111 \rangle$ nitrogen-vacancy axes oriented at $54.7^\circ$ to the surface normal. The diamond is mounted on a high-speed electric motor that is capable of rotation speeds up to $3.33$\,kHz (200,000\,rpm) in this configuration, and a purpose-built scanning-objective confocal microscope is used to optically address the NV centres. For quantum state control, we use microwave fields tuned to the ground state $m_S = 0\leftrightarrow m_S = -1$ two-level system. The timing of laser and microwave pulses for state manipulation is controlled by a pulse generator triggered by a signal synchronous with the rotation of the diamond. We denote the rotation axis as $z$, nominally orthogonal to the plane of the diamond surface. For the experiments described in this work, we typically operate with a $6.2\,$G magnetic bias field parallel to $z$. At this field strength the spin coherence time of the NV centres in the diamond is approximately $T_2\approx350\,\upmu$s (see Supplementary Material fig. S1), limited by carbon-13 nuclear spin bath dynamics. 

\subsection{Strobed Confocal Microscopy}
\begin{figure*}
	\centering
		\includegraphics{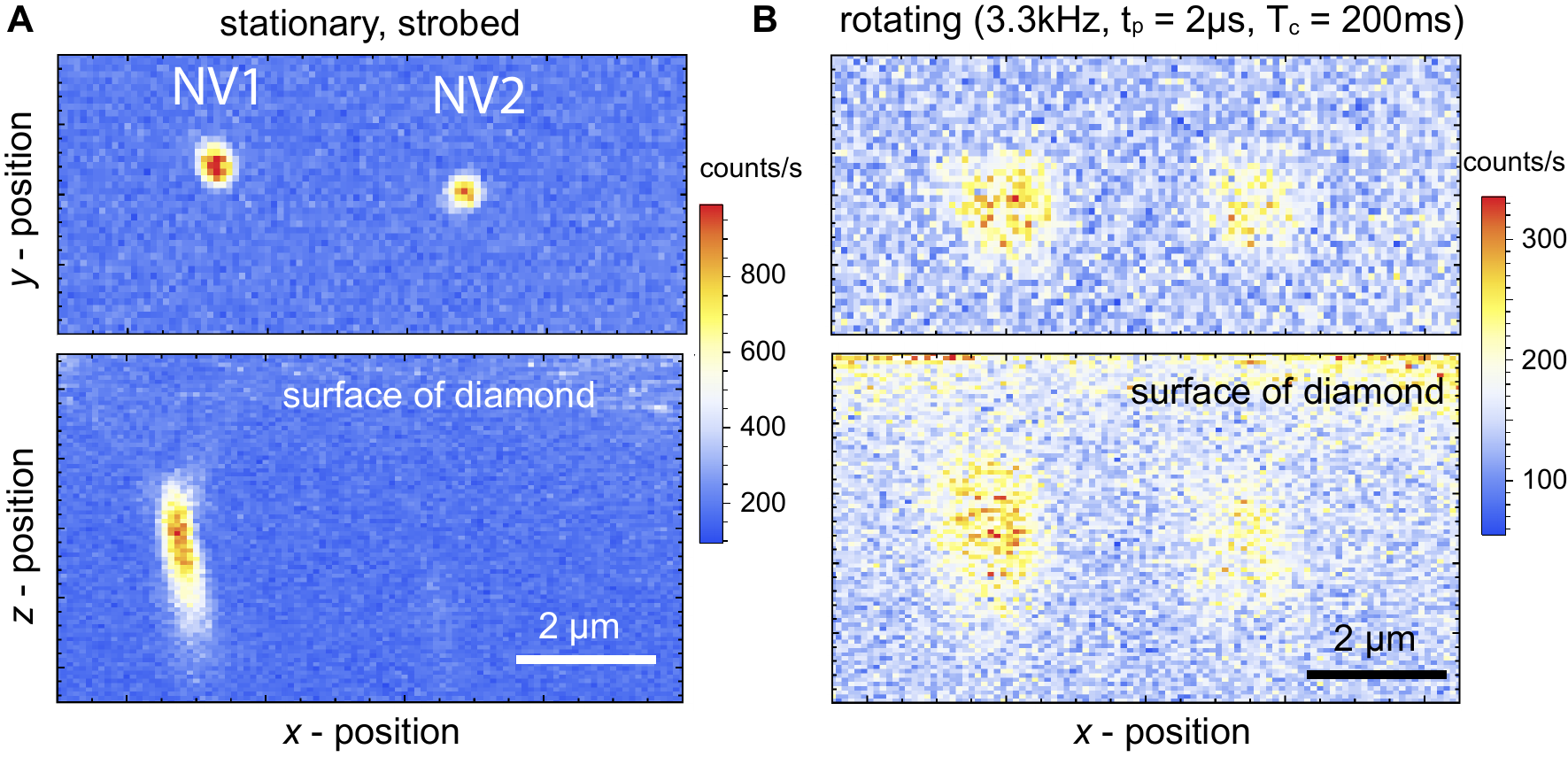}
	\caption{{\bf Strobed confocal microscopy of NV centres.} NV1 and NV2 when stationary ({\bf A}) and when rotating at high speed ({\bf B}). Upper panels are $x-y$ plane scans while lower panels are $x-z$ scans into the diamond. The color bar describes count rate at the experimental duty cycle of $D = 0.67\,\%$ (peak counts of $1\times 10^5$ recorded for $D = 1$). The stationary confocal images of NV1 and NV2 are strobed synchronously with an external $3.33\,$kHz signal from a function generator to yield an equivalent duty cycle to the rotating case. Counts are integrated for 200\,ms at each pixel with the laser pulse duration $t_\text{pulse} =2\upmu$s. ({\bf B}) The optical illumination and collection is pulsed to be synchronous with the 3.33\,kHz rotation of the diamond; blurring resulting from period jitter and wobble of the rotation is evident. NV1 was used for the remainder of the measurements described in this work.}
	\label{fig:dat}
\end{figure*}

An essential prerequisite to probing and manipulating the NV is to first locate it within the rapidly rotating diamond. Even when the host diamond is physically rotating with a period comparable to the spin coherence time of bulk diamond ($T_2 = 0.1-1$\,ms~\cite{balasubramanian_ultralong_2009}, $T_2^{-1} \sim \text{kHz}$), accessing the quantum properties of the NV under such conditions is still technically challenging. Quantum sensors must be initialized to a well defined state, subjected to quantum state manipulation gates and then read out with high fidelity after some sensing time~\cite{degen_quantum_2017}. A single NV qubit located away from the centre of rotation will execute circular motion, complicating optical addressing. We focus on two NV centres (NV1, NV2, see Fig. \ref{fig:apparatus}(A) separated by $3.6\um$ and about $r = 10\,\um$ from the centre of the diamond rotation $(r = 0)$. Fig. \ref{fig:dat} shows a confocal image created by strobing the illumination synchronously with the rotation while the position of the microscope objective is scanned. A variable time delay $t_\phi$ between the motor synchronisation signal and the laser pulse controls the instantaneous angle of the diamond and hence position (and axis orientation) of the NV. The length of the laser pulse, $t_\text{pulse}$, determines the angular displacement and thus the resulting `smear' of the NV fluorescence during imaging (see Supplementary Material fig. S2). With $f_\text{rot} = 3.33\,$kHz ($T_\text{rot} = 300\,\upmu$s) and $t_\text{pulse} = 2\,\upmu$s, NV1 subtends an angle of $2.4^\circ$ during the laser pulse. For our collection efficiency (assumed to be unchanged when rotating) and a laser power of $2.6\,$mW, we collect $N_s = 1\times 10^5$ counts/s from a stationary NV under continuous illumination; we expect to measure $N = N_s t_\text{pulse}/T_\text{rot}\sim 700$\,counts/s when rotating at $3.33\,$kHz (a duty cycle of $D = 0.67\%$). This maximum count rate is further reduced by noting the fluorescence emitted by the NV is time-dependent, since the NV samples the Gaussian intensity profile of the pumping laser beam (with $1/e^2$ diameter $d_0\approx 600\,$nm). We therefore typically measure about $350\,$ counts/s when rotating at $3.33\,$kHz. 

Figure \ref{fig:dat} shows strobed confocal images of rotating NV centres with $t_\text{pulse} = 2\,\upmu$s, compared to stationary images. The two NVs are clearly resolved and exhibit broadening due to jitter of the motor period and wobble of the motor rotation axis: the characteristic $1/e^2$ width is $\sigma = 0.9\um$, compared to stationary images where $\sigma = 0.3\um$. Timing jitter broadens the NV fluoresence azimuthally along its trajectory and wobble of the motor rotation axis broadens it primarily radially. For a laser pulse duration of $2\,\upmu$s, the azimuthal smearing is less than the width due to jitter of the rotational period ($<0.4\%$) and positional wobble of the rotation axis (see Supplementary Material fig S3). We characterized the motor rotation prior to attaching a diamond to the spindle and found the typical postitional variation of a fiducial to be $<1\um$ at $3.33\,$kHz, consistent with the blurring we observe in the strobed confocal images in Figure \ref{fig:dat}.

\subsection{Quantum state preparation, readout and control}
In a stationary experiment, the quantum state occupation probability of the NV centre is readily determined by measuring the fluorescence of the NV in some unknown superposition state relative to the fluorescence from a pure $m_S = 0 $ state. When the NV is illuminated with green light, the state contrast is as high as $30\%$ for a fully polarised $m_S = \pm1$ state and then gradually reduces in time as the NV populations are pumped into a steady state. Normalised state populations can then be extracted from a single photoluminesence time series by comparing the fluorescence just after the laser turns on, when the contrast is highest, to some time after which the NV populations have reached steady-state. Alternatively, the NV state can be determined by collecting separate fluorescence signals from the NV with and without microwave pulses applied~\cite{steiner_universal_2010}. Both of these procedures repump the NV into the $m_S = 0$ state for subsequent experimental cycles. In our experiments, the NV moves during the laser preparation and readout. As it traverses the Gaussian spatial profile of the laser beam, the NV experiences an intensity with a time dependence comparable to the optical pumping dynamics responsible for state contrast, and also emits fluorescence that is Gaussian in time. The state-dependent fluorescence of the NV that underlies state identification is therefore convolved with the time dependence of the laser intensity. 

\begin{figure}
	\centering
		\includegraphics{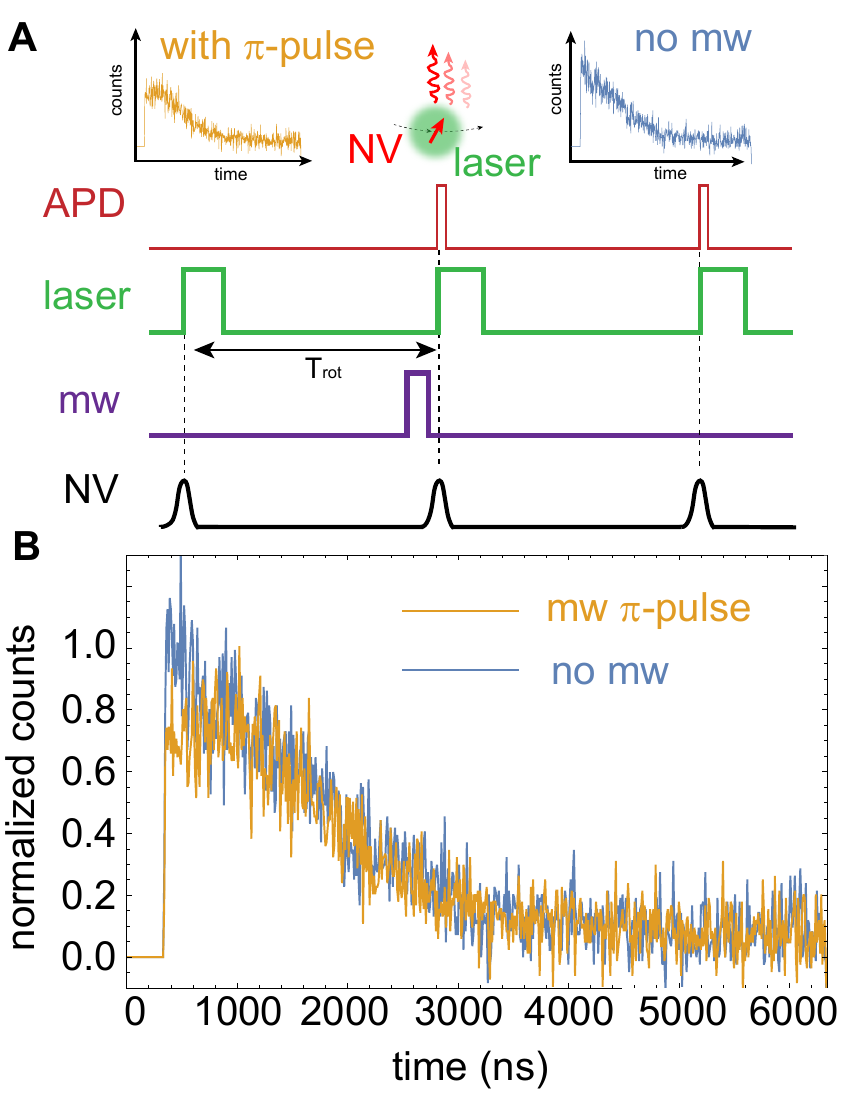}
	\caption{{\bf Photoluminesence from a rapidly rotating NV centre.} ({\bf A}) The NV spin state is determined by collecting fluorescence after the application of a microwave pulse and comparing the photon counts to the same NV when in the $m_S = 0$ bright state. ({\bf B}) The measured photoluminesence has a truncated gaussian shape in time, due to the motion of the NV through the laser beam. The initial fluorescence from the NV in the $m_S = -1$ dark state is approximately $20-30\%$ lower than after re-initialisation into the the $m_S = 0$ bright state.}
	\label{fig:PLdata}
\end{figure}

The emitted fluorescence from the moving NV is shown in Figure \ref{fig:PLdata}. A microwave pulse is applied to an NV initially polarised into $m_S = 0$, creating an unknown superposition state. We then apply green excitation light and compare the fluorescence time series from the unknown state to that collected after another rotational period with no microwave pulses applied. After the second period, the NV has been repumped to the $m_S = 0$ bright state, and the fluorescence trace can then be used to compute the normalised contrast. We routinely observe contrasts of between $20-30\%$ for resonant $\pi$-pulses, indicating that the readout pulse adequately repumps the NV back to the bright state.  

The laser can be activated or extinguished at any time during transit through the spatial extent of the laser beam, so that the NV sees a truncated Gaussian intensity profile. From a simple theoretical model~\cite{manson_nitrogen-vacancy_2006,tetienne_magnetic-field-dependent_2012} of time-dependent optical pumping of the NV, we find that the optimum time to turn on the laser and achieve maximum signal-to-noise is when the NV is almost directly under the centre of the laser beam (see Supplementary Materials fig. S4). 

In this work, we have considered the case of a tightly-focused laser beam originating from a high-NA objective, and many of the difficulties in state readout due to the transit of the NV through the beam could be mitigated by using a larger laser spot size. However, a larger laser spot size would also reduce the solid-angle collection efficiency and require more optical power. Our approach applies more generally to resolution of single centres in the case of a diamond hosting more closely-spaced NV centres. The effects of motion could also be mitigated somewhat with an NV centre positoned closer to the rotational centre.

\begin{figure}
	\centering
		\includegraphics{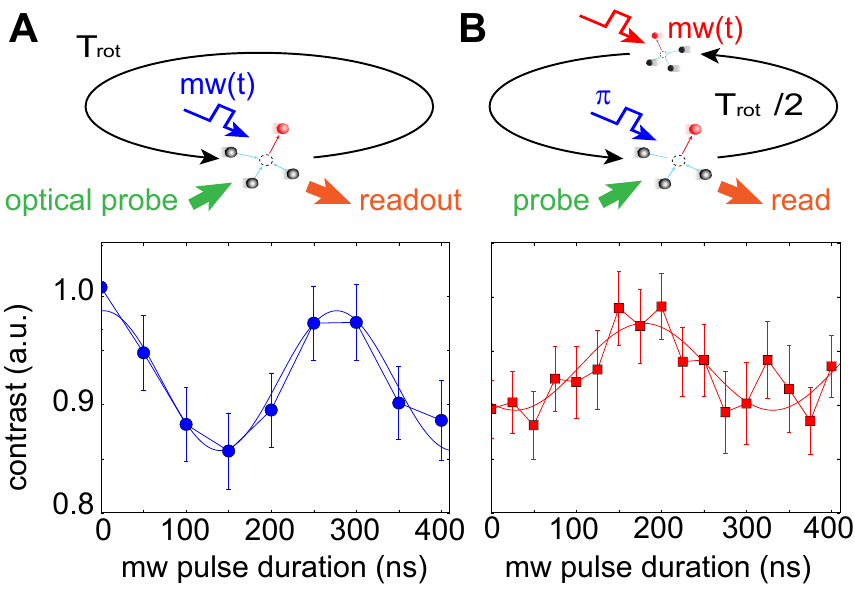}
	\caption{{\bf Quantum state control of a rapidly rotating qubit.} ({\bf A}) Varying the duration of the microwave pulse in the sequence depicted in Fig. \ref{fig:PLdata}({\bf A}) results in time-domain Rabi oscillations, which we detect by computing the normalised fluorescence. ({\bf B}) We demonstrate quantum state control of the NV spin throughout the rotation by first applying a $\pi$-pulse at $t = 0$ and then applying a variable duration microwave pulse at $t = T_\text{rot}/2$, at which the NV has moved $20\um$ away from its initial position. We then observe Rabi oscillations with the NV initialised to the $m_S = -1$ state. Error bars are standard error in computed fluoresence ratios from $>10^5$ experimental repetitions.   
}
	\label{fig:rabipholder}
\end{figure}

To demonstrate quantum state control of the rotating NV, we vary the duration of the applied resonant microwave pulse (Fig. \ref{fig:PLdata}) and measure the emitted fluorescence to determine the spin state. We observe Rabi flopping between the $m_S = 0$ and $m_S= -1$ states, as shown in Fig. \ref{fig:rabipholder}(A) with a Rabi frequency of $3.6(2)\,$MHz. We also performed measurements where we first apply a $\pi$-pulse and then a second, variable-duration microwave pulse halfway through the rotation and observe the expected population oscillations with the NV initialised to the $m_S = -1$ state. Here, the microwave pulse flips the NV spin state after $150\,\upmu$s of rotation, when it is $20\,\um$ from the laser preparation and readout region and has undergone a $180^\circ$ rotation of the NV axis. The observation of Rabi oscillations initiated from the $m_S = -1$ state in Fig. \ref{fig:rabipholder}(B) confirms the NV state can be controlled far away from the preparation and readout region.

\subsection{Spin-echo interferometry of a rotating qubit}

These results demonstrate the fundamental prerequisites to employing an NV qubit as a rotating frame quantum sensor. We now apply these state control and readouts techniques to quantum sensing in the rotating frame by performing spin-echo interferometry on the NV centre, and measure an effect unique to rotating NV centres: rotationally-induced modulation of the NV Zeeman splitting. The NV axis is not parallel to the rotation axis for our diamond, and if $B$ also makes an angle to the rotation axis the vector projection of the bias magnetic field onto the NV axis changes as the diamond rotates. The NV then experiences an effective AC (eAC) magnetic field that oscillates at the motor frequency
\begin{equation}
B_\text{eff}(t) = B_0\sin \theta_\text{NV} \sin\theta_B \cos\left(2\pi f_\text{rot} t +\phi_0\right),
\label{eq:beff}
\end{equation}
where $B_0$ is the total magnetic field strength, $f_\text{rot}$ the rotation frequency, $\phi_0$ determined by the initial phase of the rotation, $\theta_B$ the angle between the rotation axis and the magnetic field and $\theta_\text{NV}= 54.7^\circ$ the angle between the rotation axis and the NV axis. The time-dependent Zeeman shift induced by the off-axis magnetic field can be detected using spin-echo interferometry~\cite{hahn_spin_1950} conducted in the rotating frame.

Evidence of $^{13}$C nuclear spins interacting with the NV can be detected in stationary spin-echo experiments, leading to a characteristic spin-echo modulated at half the nuclear spin precession frequency~\cite{childress_coherent_2006}. In our experiments, we applied a $6.2\,$G magnetic field along the $z$-axis and measured the spin-echo signal (fig S1). Although the first $^{13}$C contrast revival ($\tau_R = 2\times 2\pi/\gamma_{13\text{C}}B_0$, $\gamma_{13\text{C}} = 1.075\,\text{kHz/G})$ coincides with one rotation period at 3.33\,kHz, when rotating the phase accumulated in a $\tau = 300\,\upmu$s spin-echo experiment due to small ($<0.01^\circ$) bias field misalignments is large enough to suppress the revival. Additionally, the presence of rotationally-induced effective magnetic fields shifts the rotating revival time away from the stationary revival time~\cite{wood_magnetic_2017}.

\begin{figure}
	\centering
		\includegraphics{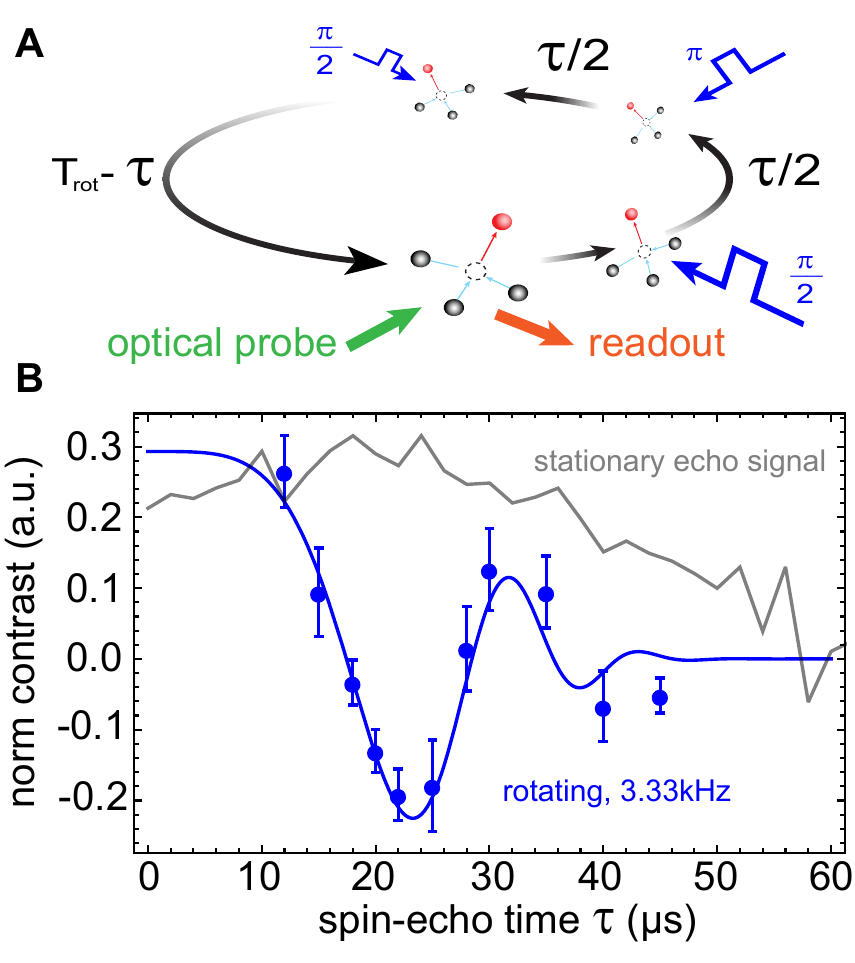}
	\caption{{\bf Quantum sensing with a single qubit in a physically rotating frame.} ({\bf A}) We examine the phase accumulated over partial rotations by performing a spin-echo experiment for interrogation times $\tau<60\,\upmu$s. Once the final $\pi/2$-pulse has projected the NV into the basis state, we wait the time remaining in the rotation for the NV to return to the preparation/readout region. ({\bf B}) Unlike the stationary echo signal (gray), we observe non-zero phase accumulation for the rotating NV. The observed fringes originate from the time-varying projection of the magnetic field onto the NV axis due to misalignment from the rotation axis (data points). The observed signal is well described by a $88\pm29$\,mG eAC field, phase-locked to the diamond rotation (fitted line). Error bars are standard error in computed fluoresence ratios averaged over three repeated experimental runs of $\sim 10^6$ repetitions.      
}
	\label{fig:echopholder}
\end{figure}

As shown in Figure \ref{fig:echopholder}(A) we consider partial rotations with the interrogation time $\tau$ short enough that the eAC field phase is slowly varying and the signal contrast has not collapsed due to the NV-$^{13}$C interaction ($\tau<60\,\upmu$s). In this region, the microwave coupling is also reasonably uniform over the spatial region considered (a $12\,\upmu$m arc). The final projective microwave pulse is then applied at the desired time $\tau$, and state readout performed after a time $T_\text{rot} - \tau$ after which the NV has completed the rotation and arrives beneath the laser. 

Figure \ref{fig:echopholder}(B) shows the spin-echo signal for a single NV rotating at $3.33\,$kHz, demonstrating the convergence of the quantum measurement and control protocols in the rotating frame described in this work. In comparison with the stationary echo signal, we observe fringes due to the presence of eAC fields. The spin-echo signal observed is consistent with the NV experiencing an eAC field (Eq. \ref{eq:beff}) of magnitude $B_0\sin\theta_\text{NV}\sin\theta_B=88\pm29$\,mG, which corresponds to $\theta_B = 1.0\pm0.3^\circ$. The uncertainty in the inferred field derives from the highly covariant nature of the parameters used to model the spin-echo signal in Fig.\ref{fig:echopholder} and is not representative of the ultimate sensitivity of the technique. Modulating the NV Zeeman shift by rotation could, in principle, offer a simple alternative to existing schemes~\cite{hong_coherent_2012, ajoy_dc_2016} of surpassing $T_2^*$-limited DC field sensitivities, pushing the relevant sensing time to $T_2$.  

\section{Discussion}

Our results establish the NV centre as a tool to study the effects of physical rotation on individual quantum spins. Such effects include rotationally-induced strain, geometric phase~\cite{maclaurin_measurable_2012, kowarsky_non-abelian_2014} and rotationally-induced magnetic fields~\cite{barnett_gyromagnetic_1935, chudo_observation_2014,chudo_rotational_2015,wood_magnetic_2017}, all important considerations for gyroscopic applications with NVs~\cite{ledbetter_gyroscopes_2012, ajoy_stable_2012}. Rotation is typically an issue for motional sensors, such as nanodiamonds within biological cells~\cite{mcguinness_quantum_2011} and a better understanding of the effect of rotation on single qubits could lead to more incisive quantum sensing in living organisms~\cite{maclaurin_nanoscale_2013}. 

The quantum state manipulations in the physically rotating frame demonstrated here form the basis of new additions to the quantum sensing toolbox. Further extension to more complicated pulse sequences, with a concomitant increase in the incisiveness of the measurement, is therefore possible. A more direct investigation of the spin dynamics between a coupled NV-nuclear spin system, using correlation spectroscopy~\cite{laraoui_high-resolution_2013}, for example, could investigate with a single spin sensor the low to zero-field transition between competing decoherence mechanisms in the nuclear spin bath~\cite{wood_magnetic_2017}.

Rotation could potentially improve the coherence, and hence precision, of quantum sensors. Magic-angle spinning is used in NMR experiments to narrow transition linewidths: rotation of a solid-state nuclear spin sample with $B$ oriented at $54.7^\circ$ emulates motional averaging of internuclear magnetic dipole-dipole interactions in liquids~\cite{andrew_magic_1981}. Considerable effort has been aimed at implementing nanoscale NMR sensing with the NV centre~\cite{staudacher_nuclear_2013,sushkov_magnetic_2014,lovchinsky_nuclear_2016} and motional narrowing of proton NMR signals in liquids has been observed with NV centres~\cite{staudacher_probing_2015}. It would be interesting to explore the effect of rotation at the magic angle on the internuclear interactions, either between NMR target spins in solids or between nuclear spins (such as $^{13}$C) within the diamond, the latter potentially increasing the coherence time of the NV centre itself.

\section{Methods and Materials}
{\bf The NV centre in diamond} In the absence of magnetic fields, the electron spin of the NV centre is quantized along the nitrogen-vacancy axis and forms a ground state spin-triplet system: degenerate $m_S = \pm1$ states and a $m_S = 0$ state separated by the zero-field splitting $D_\text{zfs} = 2.87\,\text{GHz}$. Spin-conserving optical transitions from the ground state to an excited triplet state $1.945\,$eV higher in energy are efficiently driven with $532\,\text{nm}$ light. The NV emits red photons at the $638\,$nm zero-phonon line and in a broad vibrational sideband. Peak emission from NVs in the $m_S = 0$ ground state is up to $30\,\%$ brighter than that from $m_S = \pm1$ states due to an intersystem crossing that results in non-radiative relaxation from the $m_S =\pm1$ states \cite{manson_nitrogen-vacancy_2006}. The spin state of the NV can therefore be determined via the fluorescence contrast between the bright $m_S=0$ and dark $m_S =\pm1$ states \cite{steiner_universal_2010}.  
  
Our diamond sample is a high-purity single-crystal CVD diamond cut along the $(100)$ plane, with a $50\,\mu$m layer of $99.8\% \,^{12}$C grown on the surface. The concentration of nitrogen impurities is $<10$\,ppb. A 0.95 NA microscope objective (Olympus UMPlanFl) mounted on a three-axis piezoelectric stage (PI 611.3S) focuses 532\,nm laser light to a $600$\,nm spot size onto the diamond, which is mounted on a high-speed electric motor (Celeroton CM-2-500). Red fluorescence from NV emitters is collected by the same lens and directed onto an avalanche photodiode (Perkin-Elmer SPCM-15-14), in a confocal microscope configuration. Photon counts from the APD are time-tagged and processed by a field-programmable gate array (FPGA, Opal Kelly XEM3005) before computer analysis with custom Python code. 

 A 20\,$\upmu$m diameter copper wire located $100\,\upmu$m above the surface of the diamond is used to produce microwave magnetic fields for state manipulation. Current-carrying coils aligned along the ${x,y,z}$ axes are used to apply a bias field field along the $z$ axis, with the $x, y$ coils used to eliminate off-axis field components. The coils are driven by a Hameg HMP4030 power supply. Control pulses synchronous with the rotation of the diamond are provided by an Opal Kelly XEM6310 FPGA triggered by the coil-current phase synchronisation signal from the motor converter (Celeroton CC-75-500). The FPGA pulses the microwaves (produced by a Windfreak SynthHD) and laser by actuating high-speed microwave switches (Minicircuits ZASWA-2-50DR).   

{\bf Spin state readout}
We used high laser powers of up to 3\,mW within a narrow laser spot size ($I_0\approx 1\,\text{MW\,cm}^{-2}$), well above the saturation intensity. We inferred the ability of the readout laser pulse to subsequently repump the NV to the bright state by comparing the fluoresence after readout for different states. Regardless of what state the NV was in before the first readout laser pulse, we observed the same photoluminesence trace with the second readout pulse.  

{\bf Pulse fidelity during rotation}  The $[111]$-oriented NV axis $\boldsymbol{n}$ is always $54.7^\circ$ to the rotation axis in our experiment since our diamond is mounted on its $(100)$ face. Unless the applied magnetic bias and microwave fields required for state manipulation are also parallel to the rotation axis, the Zeeman shift ($\propto \boldsymbol{n}\cdot\boldsymbol{B}$) and microwave coupling ($\propto |\boldsymbol{n}\times\boldsymbol{B}_\text{mw}|$) vary as the diamond rotates. Microwave pulses that interrogate the NV throughout the rotation can thus be off resonant (in the case of large field misalignments) or drive imperfect spin rotations when the Rabi frequency varies substantially. We align the magnetic field to the rotation axis to within $1^\circ$, which results in a uniform detuning throughout the rotation. For sequences using multiple microwave pulses, such as spin-echo, we determine the Rabi frequency at each point in the rotation and calibrate the requisite pulse durations to achieve the desired spin rotations.

%

\section*{Acknowledgments}
We acknowledge valuable discussions with D. A. Simpson, A. D. Stacey and F. Jelezko. We thank J.-P. Tetienne for assistance with simulating time-dependent optical pumping in the NV. A. M. M. would like to thank the Institute of Advanced Study (Durham University, U.K.) for hosting him during the preparation of this manuscript. This work was supported by the Australian Research Council Discovery Scheme (DP150101704). 

\end{document}